\documentclass[epj,referee]{svjour}
\usepackage{graphics}
\begin{document}
\title{Study of forbidden atomic transitions on $D_2$  line using Rb nano-cell placed in external magnetic field}
\author{G. Hakhumyan\inst{1,2} \and C. Leroy\inst{2}\and R. Mirzoyan\inst{1,2} \and Y. Pashayan-Leroy\inst{2} \and D. Sarkisyan\inst{1}
}                     % Do not remove
\institute{Institute for Physical Research, NAS of Armenia, Ashtarak 0203, Armenia \and
Laboratoire Interdisciplinaire Carnot de Bourgogne, UMR CNRS 5209, Universit\'{e} de Bourgogne, 21078 Dijon Cedex, France}
\date{Received: date / Revised version: date}
\abstract{
By experimental exploration of the so-called $\lambda $-Zeeman technique based
on Rb nano-cell use we reveal for the first time a strong modification of the
probability of the $^{87}$Rb, $D_2$ line $F_g=1 \rightarrow F_e=0, 1, 2, 3$ atomic transitions, including
forbidden $F_g=1, m_F =0 \rightarrow F_e=1, m_F =0$ and $F_g=1, m_F =-1 \rightarrow F_e=3, m_F =-1$ transitions
(these are forbidden transitions when $B = 0$) in a strong external magnetic
field $B$ in the range of $100 - 1100$~G. For $\pi$-polarized exciting diode laser
radiation ($\lambda  = 780$~nm) these forbidden transitions at $B > 150$~G are
among the strongest atomic transitions in the detected transmission spectra.
Frequency shifts of the individual hyperfine transitions versus magnetic field
are also presented: particularly, $F_g=1, m_F =+1 \rightarrow F_e=1, m_F =+1$ atomic transition has a
unique behavior, since its frequency remains practically unchanged when $B$
varies from $100$ to $1100$~G. Developed theoretical model well describes the experiment.
} %end of abstract
\authorrunning {G. Hakhumyan, C. Leroy}
\titlerunning {Forbidden $D_2$ line in Rb nano-cell  in $B$-field}
\maketitle
\section{Introduction}
\label{intro}
Alkali atoms, particularly Rb atoms, are widely used in laser atomic physics due to strong
atomic transitions with the wavelength located in the near infrared region.  It is also
important that there are available diode lasers with good parameters which wavelength
is resonant with the atomic transitions. Rb atoms are widely used in laser cooling
experiments, information storage, spectroscopy, magnetometry \textit{etc} \cite{Budker,Mechede}.
Of a special interest are the $^{87}$Rb atoms used in Bose-Einstein Condensates (BEC)
experiments \cite{Budker}. That's why a detailed knowledge of the behavior of Rb atomic transitions,
particularly, in an external magnetic field is of high importance. It is well-known
that atomic energy levels split in a magnetic field into Zeeman sub-levels, and frequency shifts of atomic transitions between ground and upper sub-levels (optical domain)
deviate from the linear behavior in quite moderate magnetic field \cite{Tremblay,Weis}. Also,
atomic transition probabilities undergo significant changes  depending
on external magnetic $B$-field \cite{Tremblay}. Usually, frequency separation
between atomic transitions in an external magnetic field of $50 - 1000$~G
achieves $20 - 200$~MHz. However, because of Doppler broadening ($\sim 500$~MHz),
in order to study separately each individual atomic transition behavior one should implement a
technique providing sub-Doppler resolution.\\
\indent It is known that, with the saturated absorption (SA) technique, the sub-Doppler spectral
resolution can be achieved using conventional centimeter-scale cells. In Refs.~\cite{Momeen,Ban}
the SA technique is used to study spectra of $D_2$ line of Rb atoms. However, one of the
significant disadvantages of the SA technique is the presence of
the so-called cross-over resonances in the spectra. In a magnetic field, these resonances
split into numerous components, making the spectrum very difficult to analyze.
This restricts the magnitude of acceptable magnetic field below $100$~G. Another
disadvantage of the SA technique is the fact that, the amplitudes of velocity selective optical pumping
(VSOP) resonances formed in SA spectrum do not correspond to the probabilities
of the corresponding atomic transitions. This additionally strongly complicates the
analysis of spectra. Note, that sub-Doppler spectral resolution could be obtained
 by using expensive and complicated systems based either on cold and trapped atoms
 or with the help of collimated several-meter long Rb atomic beam propagated in vacuum conditions.\\
\indent Recently, it has been demonstrated that a one-dimensional nano-metric thin
cell (NTC) filled with Rb atoms is a very convenient tool to obtain sub-Doppler
spectral resolution when the thickness $L$ of atomic vapor column is either
$L = \lambda/2$ or $L = \lambda$, where $\lambda$ is the laser radiation wavelength
resonant with the Rb $D_1$ or $D_2$ lines ($\lambda=794$~nm or $780$~nm). In case
of the thickness  $L = \lambda/2$ it is more convenient to use the fluorescence of the NTC since
the spectrum linewidth is $7-8$ times narrower than that of the Doppler width obtained with a
conventional cm-size Rb cell. The method is called "half-$\lambda$ Zeeman technique" (HLZT)~\cite{Sarkis1,Sarkis2,Hakhum1}.\\
\indent In case of the thickness $L = \lambda$ spectrally-narrow VSOP resonances appear
at laser intensities $\sim 10$~mW/cm$^2$ in the transmission spectrum of NTC. The formation of VSOP
resonances with the help of NTC has several advantages in comparison
with the SA technique: i) the absence of cross-over resonances, which is very important
for some applications, particularly, when an external magnetic field is applied; ii)
the ratio of amplitudes of VSOP resonances is close to the ratio of the corresponding
atomic transition probabilities; iii) a single beam transmission is used; iv) the laser
power required for the formation of VSOP
resonances is as low as $0.1$~mW. In a magnetic field
these VSOP resonances are split into several new components, the number of which depends on the
quantum numbers $F$ of the lower and upper levels, while the amplitudes
and frequency positions of the components depend on $B$-field. This method allows one to study
separately each individual atomic transition behavior ("$\lambda$-Zeeman
technique" (LZT))~\cite{Hakhum2,Hakhum3,Leroy,Sarkis3}.\\
\indent Below are presented the results of the experimental study of the $^{87}$Rb, $D_2$ line
$F_g=1 \rightarrow F_e=0, 1, 2, 3$ atomic transitions, including forbidden $F_g=1, m_F =0 \rightarrow F_e=1, m_F =0$
and $F_g=1, m_F =-1 \rightarrow F_e=3, m_F =-1$ transitions for $\pi$-polarized exciting laser radiation.
For this study Rb NTC is placed in a strong external magnetic
field $B$ varying in the range of $100 - 1100$~G. A theoretical model applied to
describe the experimental results is presented.
\section{Experiment}
\label{sec:Exp}
%and \cite{RefJ}
\subsection{Experimental setup}
\label{subsec:SetUp}
\begin{figure}
\begin{center}
\resizebox{0.6\columnwidth}{!}
{
\includegraphics{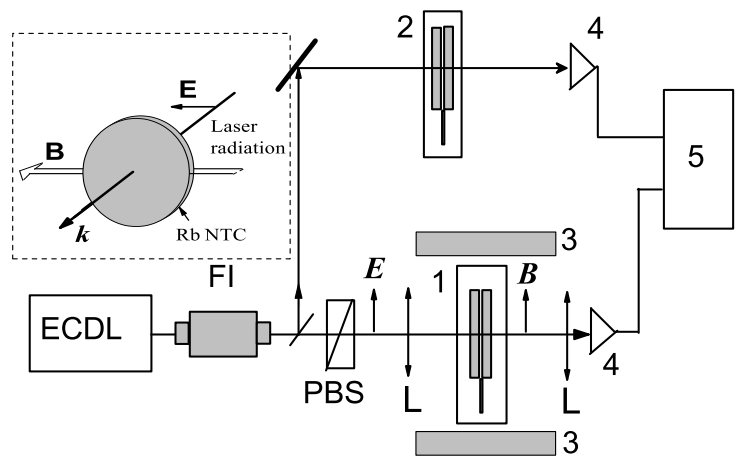}
}
\caption{Sketch of the experimental setup.
 \textit{ECDL} - diode laser, \textit{FI} - Faraday isolator,  \textit{1} - NTC in the oven, \textit{2} - an auxiliary NTC and the oven, \textit{PBS} - polarization beam splitter, \textit{3} - permanent magnets (PMs), $L$ - lenses, \textit{4} - photodetectors, \textit{5} - digital storage oscilloscope, \textbf{E} - electrical field of laser radiation, \textbf{B} - magnetic field applied to NTC, $\textbf{B}// \textbf{E}$.
  Configuration of the magnetic measurement is presented in the inset in the dashed square.}
\label{fig:SetUp}
\end{center}
\end{figure}
The first design of the NTC (also called extremely thin cell) consisting
of windows and a vertical side arm (a metal reservoir), is presented in \cite{Sarkis4}.
Later, this design has been somehow modified and a typical example of a recent
version is given in \cite{Hakhum1}. The used NTC has garnet windows of $2.4$~mm thickness.
The NTC is filled with a natural mixture of $^{85}$Rb ($72.2 \%$) and $^{87}$Rb ($27.8\%$).
Our study concerns the region of $L \approx \lambda \approx 780$~nm. The temperature limit of the NTC
operation is 400 ${}^\circ$C. The NTC operated with a specially designed oven (made from
non-magnetic materials) with two ports for laser beam transmission. The source
temperature of the atoms of the NTC was 120 ${}^\circ$C, corresponding to the vapor
density $N = 2\times10^{13}$~cm$^{-3}$, but the windows were maintained at a temperature that was 20~${}^\circ$C higher.\\
\indent The sketch of the experimental setup is shown in Fig. \ref{fig:SetUp}. The $\pi$-polarized
beam of extended cavity diode laser (ECDL,
$\lambda=780$~nm, $P_L=30$~mW, $\gamma_L< 1$~MHz)
resonant with the $^{87}$Rb, $D_2$ transition frequency, is focused (the laser spot diameter
is $\leq  0.1$~mm) at nearly normal incidence onto the Rb NTC \textit{1} with the vapor
column thickness $L = \lambda = 780$~nm. To avoid feedback a Faraday insulator is applied.
A polarization beam splitter \textit{PBS } is used to purify initial linear radiation
polarization of the laser. A part of the pumping radiation was directed to the auxiliary
(reference) Rb NTC \textit{2}, which was in zero magnetic field; transmission spectrum of
 this NTC is used as a frequency reference. The transmission signal was detected by a
 photodiode \textit{4} and was recorded by Tektronix TDS $2014$~B digital four-channel storage
 oscilloscope \textit{5}. Moderate magnetic fields in the range of $10 - 200$~ G are produced
 by Helmholtz coils, while in order to produce strong magnetic fields ($B > 200$~G)
 two strong permanent magnets (PMs) \textit{3} are used \cite{Hakhum3}. In both cases \textbf{B}
 is directed along the laser electric field direction \textbf{E}
 ($\textbf{B} // \textbf{E}$). The configuration of the magnetic measurement
 is presented in the inset in Fig.~\ref{fig:SetUp}. The $B$-field strength was measured by a calibrated Hall gauge.
\subsection{ Experimental results and discussion}
\label{subsec:ExpResults}
\begin{figure}
\begin{center}
\resizebox{0.5\columnwidth}{!}
{
\includegraphics{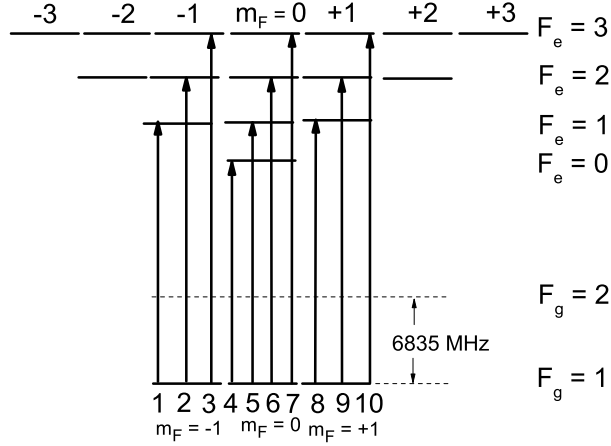}
}
\caption{Energy levels diagram of the $D_2$ line of $^{87}$Rb and $F_g=1 \rightarrow F_e=3$ atomic transitions
for $\pi$-polarized exciting laser radiation ($\Delta m_F = 0$)}
\label{fig:levels}
\end{center}
\end{figure}

\begin{figure}
\begin{center}
\resizebox{0.6\columnwidth}{!}
{
\includegraphics{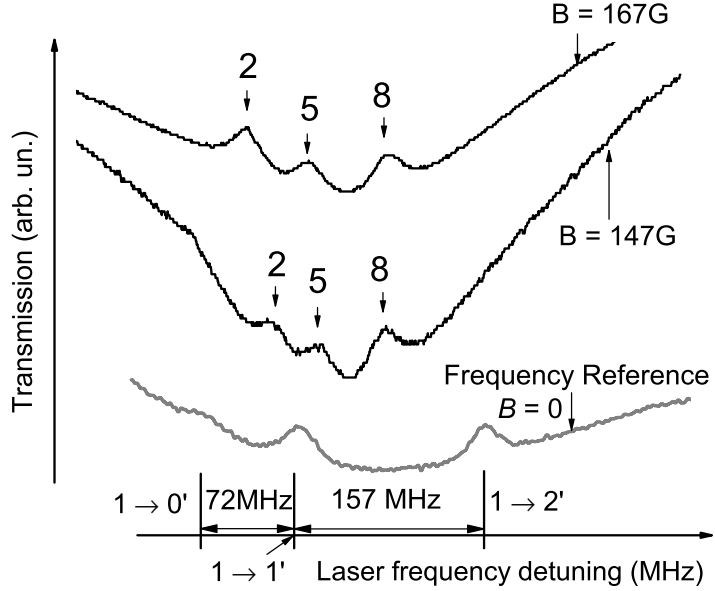}
}
\caption{Transmission spectra from the Rb NTC ($L = \lambda = 780$~nm) for transitions
$F_g=1 \rightarrow F_e=0, 1, 2, 3$ vs magnetic field. VSOPs numbers denote the corresponding
transitions depicted in Fig.~\ref{fig:levels}. Forbidden transition $F_g=1, m_F =0 \rightarrow F_e=1, m_F =0 $
labeled \textit{5} at $B \sim 150$~G is among the three strongest atomic transitions. The
lower grey curve is the transmission spectrum from the reference Rb NTC ($L = \lambda$).
The spectra are shifted vertically for convenience.}
\label{fig:Lambda}
\end{center}
\end{figure}

The atomic transitions $F_g=1 \rightarrow F_e =0, 1, 2, 3$ between magnetic sub-levels of hyperfine
states for the $^{87}$Rb, $D_2$ line (optical domain) in the case of $\pi$-polarized laser
radiation excitation are depicted in Fig.~\ref{fig:levels}. Note that when  $B=0$ according to the
selection rules the  atomic transitions with
the corresponding $\Delta F = 2$, namely $F_g=1 \rightarrow F_e=3$ and
 $F_g=1, m_F =0 \rightarrow F_e=1, m_F =0$ transitions are strongly forbidden, while all other presented
 transitions with $\Delta F = F_g - F_e = 0$, $\pm$ 1 and $\Delta m_F = 0$  are allowed \cite{Demtroder}.\\
\indent As it was recently shown, $\lambda$-Zeeman technique implemented in case
of $\sigma^{+}$ (left circular) polarized excitation allows one to study separately each
individual atomic transition behavior in an external magnetic field \cite{Hakhum2,Hakhum3}.
As demonstrated below, the $\lambda $-Zeeman technique implemented in case
of $\pi$-polarized excitation is also very convenient, since the examination of the VSOP
resonances formed in the NTC allows one to obtain, identify, and investigate each
individual atomic transition between the Zeeman sub-levels in the transmission
spectrum of the $^{87}$Rb $D_{2}$ line in a very wide range of magnetic fields
from a few tens up to several thousands of Gauss.\\
\indent The two upper curves in Fig.~\ref{fig:Lambda} show the transmission spectra from
the Rb nano-cell $L = \lambda = 780$~nm for the magnetic field $B \approx 147$~G and $167$~G,
with $\pi$-polarized exciting laser radiation (VSOPs numbers denote the corresponding transitions
depicted in Fig.~\ref{fig:levels}). The magnetic field is produced by Helmholtz coils (the maximum
available $B$-field is $ \sim 200$~G). The splitting and shifts of the three VSOP
resonances \textit{2}, \textit{5}, and \textit{8} are clearly seen  in Fig.~\ref{fig:Lambda}.
Thus, the remarkable
result is that the forbidden transition $F_g=1, m_F =0 \rightarrow F_e=1, m_F =0$ labeled \textit{5}
at $ B \sim 150$~G is among the three strongest atomic transitions, while the other transitions
 have smaller probabilities, and thus are not detectable in the spectra (this is confirmed
 by the theory - see below). The lower grey curve is the transmission spectrum as given by the
 reference nano-cell which shows the positions of the atomic transitions, i.e., the
 VSOP resonances (with the linewidth of $\sim 20$~MHz) for $ B \sim 0$. We measure the
 atomic frequency shifts with respect to the initial position
 of $F_g=1 \rightarrow F_e=1, 2, 3$ transition
  in the studied nano-cell in the magnetic field $\textbf{B}$ directed along the laser radiation electric
  field $\textbf{E}$.\\
\begin{figure}
\begin{center}
\resizebox{0.6\columnwidth}{!}
{
\includegraphics{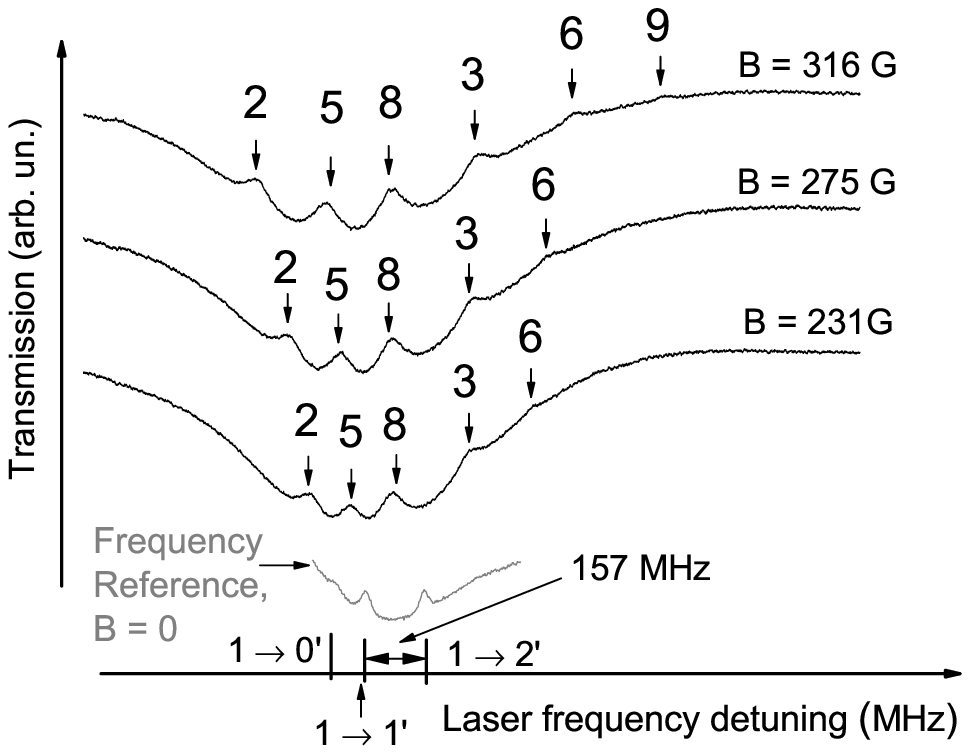}
}
\caption{Transmission spectra from the Rb NTC $L = \lambda = 780$~nm
for atomic transitions $F_g=1 \rightarrow F_e=0, 1, 2, 3$ versus magnetic field $B$: 231, 275 and 316 G.
Numbers denote the corresponding transitions depicted in Fig.~\ref{fig:levels}.
Forbidden transition $F_g=1, m_F =-1 \rightarrow F_e=3, m_F =-1$ labeled \textit{3} at $B \geq  230$~G
is among the four strongest atomic transitions. The lower grey curve
is the transmission spectrum from the reference Rb NTC $L = \lambda = 780$~nm.}
\label{fig:Lambda2}
\end{center}
\end{figure}
\indent Figure~\ref{fig:Lambda2} demonstrates the transmission spectra for the atomic transitions
$F_g=1 \rightarrow F_e=0, 1, 2, 3$ at the following values of the magnetic field $B$: 231, 275 and 316 G.
As mentioned, a strong magnetic field is produced by two PMs (with the diameter of $60$~mm)
placed on the opposite sides of the nano-cell oven and separated by a variable distance.
To control the magnetic field value, one of the magnets is mounted on a micrometric
translation stage for longitudinal displacement. The splitting and shifts of the six VSOP
resonances \textit{2}, \textit{5}, \textit{8}, \textit{3}, \textit{6} and \textit{9} are
clearly seen. Note, that at $B \geq  230$~G the forbidden transition labeled \textit{5}
together with another forbidden transition
$F_g=1, m_F =-1 \rightarrow F_e=3, m_F =-1$ labeled \textit{3}  is among the four
strongest atomic transitions (this is also confirmed by the theory - see below).\\
\begin{figure}
\begin{center}
\resizebox{0.6\columnwidth}{!}
{
\includegraphics{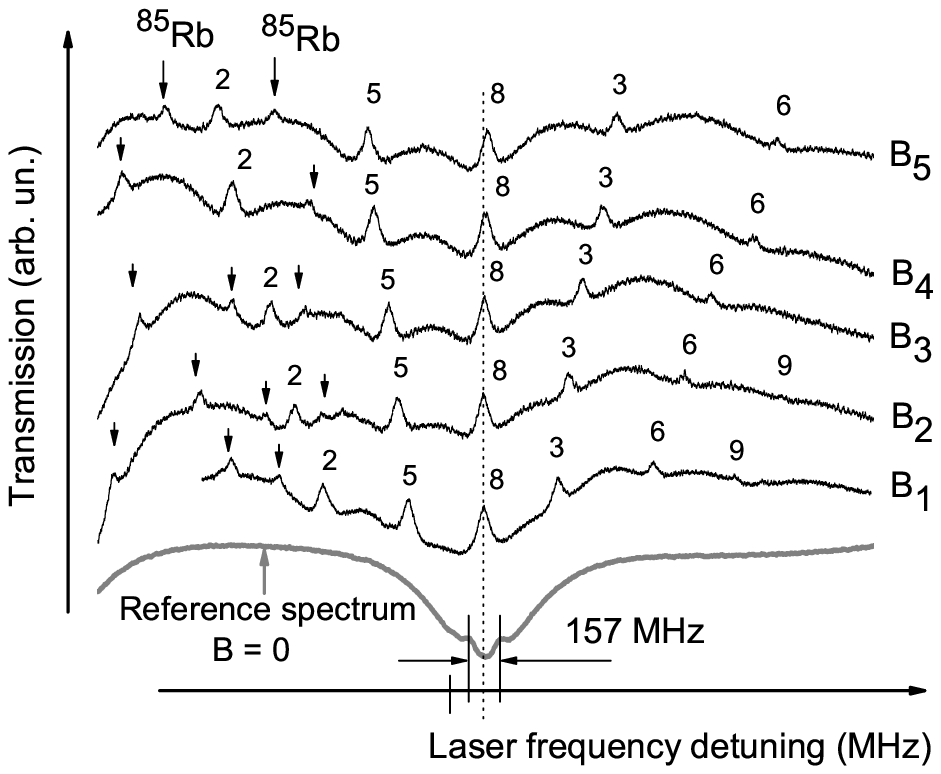}
}
\caption{Transmission spectra from the Rb NTC $L = 780$~nm for atomic transitions
$F_g=1 \rightarrow F_e=0, 1, 2, 3$ versus magnetic field  $B_{1, 2, 3, 4, 5}  = 700, 740, 865, 1010$
 and $1080$~G, correspondingly. Numbers denote the corresponding transitions depicted in Fig.~\ref{fig:levels}.
Forbidden transitions \textit{5} and \textit{3} at $B \geq  700$~G  are among the four strongest
atomic transitions. The atomic transition labeled \textit{8} is a unique one since its frequency
remains practically unchanged (vertical dotted line is presented for the eye-guide)
when $B$ varying in the range of $100 - 1100$~G. The lower grey curve is the reference one.}
\label{fig:Lambda3}
\end{center}
\end{figure}
\indent Figure~\ref{fig:Lambda3} shows transmission spectra for the atomic transitions
for the following values of the magnetic fields, $B_{1, 2, 3, 4, 5}  = 700, 740, 865, 1010$ and $1080$~G,
correspondingly. Again, the splitting and shifts of the six VSOP resonances \textit{2}, \textit{5},
\textit{8}, \textit{3}, \textit{6}
and \textit{9} are clearly seen. Importantly, the both forbidden transitions labeled \textit{3} and \textit{5}
are always among the four strongest atomic transitions. Note, that at $B \geq  700$~G
the $^{85}$Rb, $D_2$ line, $F_g=2 \rightarrow F_g=1, 2, 3$ atomic transitions (marked
 by the vertical arrows) are also
 detected. However, up to $\sim 1100$~G the influence of the $^{85}$Rb,
 $D_2$ line atomic transitions doesn't affect strongly the $^{87}$Rb spectra. A striking
 point is also as follows: the variation of the magnetic field in the range of
 $100 - 1100$~G practically doesn't cause frequency shift of
 unique $F_g=1, m_F =+1 \rightarrow F_e=1, m_F =+1$ atomic transition labeled \textit{8}
 (this is also confirmed by the theory - see here after).
\section{ Theoretical model and discussions}
\label{Theory}
\begin{figure}
\begin{center}
\resizebox{0.6\columnwidth}{!}
{
\includegraphics{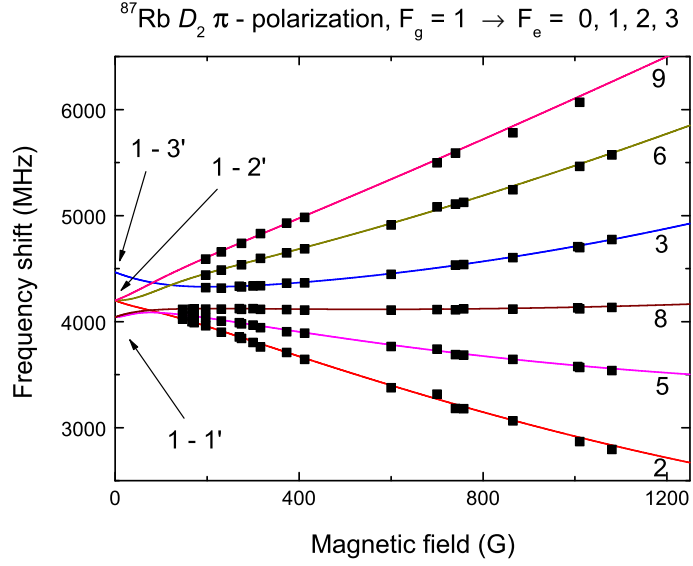}
}
\caption{Frequency shift of components \textit{2}, \textit{5}, \textit{8}, \textit{3}, \textit{6} and \textit{9} versus $B$-field
relative to the initial position (indicated by the arrows at $B = 0$),
the black squares are the experimental results and the solid curves are the
calculated ones (numbers denote the corresponding transitions, see Fig.~\ref{fig:levels})}
\label{fig:FrShift}
\end{center}
\end{figure}

In this work the main interest is to study the behavior of the $^{87}$Rb, $F_g =1 \rightarrow F_e=0, 1, 2, 3$
transitions, $D_2$ line in the case of $\pi $-excitation. Theoretical model describes how to provide
the calculations of separated transitions' frequencies and amplitude modification undergo
external magnetic field (the details of the theory are presented in \cite{Tremblay,Auzinsh}).
We adopt a matrix representation in the coupled basis, that is, the basis of the
unperturbated atomic state vectors $\left|\left.(n=5),\ L,\ J,\ F,\ m_F\right\rangle \right.$
to evaluate the matrix elements of the Hamiltonian describing our system. In this basis,
the diagonal matrix elements are given by
\begin{equation}
\label{eq:MatrixElements}
\left\langle F,m_F|H|F,m_F\right\rangle = E_0(F) + \mu_B g_F m_F B,
\end{equation}
where $E_0\left(F\right)$ is the initial energy of the sub-level
$\left|(n=5), L, J, F, m_{F}\right\rangle \equiv \left|F, m_F\right\rangle$
%$\left|\left.\left(n=5\right),L,J,F,m_F \right \rangle \right. \equiv \left | \left.F,m_F \right \rangle \right.$
and $g_F$ is the effective Land\'{e} factor.\\
\indent The off-diagonal matrix elements are non-zero for levels verifying the selection rules
$\Delta L = 0, \Delta J = 0, \Delta F = \pm 1, \Delta m_F = 0$,
\begin{equation}
\label{eq:SelRules}
\begin{array}{r}
\left\langle F-1, m_F|H|F, m_F\right\rangle = \left\langle F, m_F|H|F-1, m_F\right\rangle = -\frac{\mu_BB}{2}(g_J-g_I)\\
\times \left(\frac{[(J+I+1)^2-F^2][F^2-(J-I)^2]}{F}\right)^{1/2}\left(\frac{F^2-m_F^2}{F(2F+1)(2F-1)}\right)^{1/2}.
\end{array}
\end{equation}
The diagonalization of the Hamiltonian matrix allows one to find the
eigenvectors and the eigenvalues, that is to determine the eigenvalues
corresponding to the energies of  Zeeman sub-levels and the new states
vectors which can be expressed in terms of the initial unperturbed atomic state vectors,
\begin{equation}
\label{eq:Eig1}
\left|\Psi(F_e, m_{e})\right\rangle = \sum_{F_e^{\prime}} \alpha _{F_eF_e^{\prime}}^e(B)|F_e^{\prime}, m_{e}\rangle
\end{equation}
and
\begin{equation}
\label{eq:Eig2}
\left|\Psi(F_g, m_{g})\right\rangle = \sum_{F_g^{\prime}}\alpha _{F_gF_g^{\prime}}^g(B)|F_g^{\prime}, m_{g}\rangle.
\end{equation}
The state vectors $\left|F_e^{\prime}, m_e\right\rangle$ and $\left|F_g^{\prime}, m_g\right\rangle$
are the unperturbated state vectors, respectively, for the excited and the ground states.
The coefficients $\alpha _{F_eF_e^{'}}^e(B)$ and $\alpha _{F_gF_g^{'}}^g(B)$
are mixing coefficients, respectively, for the excited and the ground states;
they depend on the field strength and magnetic quantum numbers $m_g$ or $m_e$.
Diagonalization of the Hamiltonian matrix for $^{87}$Rb, $D_2$ line, in case of
$\pi $-polarization of exciting radiation, allows obtaining the shift
of position of energy levels in presence of external magnetic field.\\
\indent The probability of a transition, induced by the interaction of the
atomic electric dipole and the oscillating laser electric field is
proportional to the spontaneous emission rate of the associated
transition $A_{eg}$, that is, to the square of the transfer coefficients
modified by the presence of the magnetic field
\begin{equation}
\label{eq:Prob}
W_{eg}\propto A_{eg} \propto a^2\left[\Psi (F_e, m_{e}); \Psi(F_g, m_{g});q\right],
\end{equation}
The transfer coefficients are expressed as
\begin{equation}
\label{eq:Coef}
a\left[\Psi(F_e, m_{e}); \Psi(F_g, m_{g});q\right] =
\sum_{F_e^{'} F_g^{'}} \alpha_{F_eF_e^{'}}^e (B)a\left( \Psi(F_e, m_{e}); \Psi(F_g, m_{g}); q \right) \alpha_{F_gF_g^{'}}^g (B),
\end{equation}
where the unperturbated transfer coefficients have the following definition
\begin{equation}
\label{eq:Unpert}
\begin{array}{l} {a\left(\Psi(F_e, m_{e}); \Psi(F_g, m_{g}); q\right) = (-1)^{1 + I + J_e + F_e + F_g - m_{e} } } \\
 {\times \sqrt{2J_e + 1} \sqrt{2F_e + 1} \sqrt{2F_g + 1} \left(
 \begin{array}{ccc}
 F_e      & 1 & F_g \\
 -m_{e} & q & m_{g}
 \end{array}\right)\left\{
 \begin{array}{ccc}
 F_e & 1 & F_g \\
 J_g & I & J_e
 \end{array}\right\}}
\end{array},
\end{equation}
the parenthesis and curly brackets denote, respectively, the $3j$ and $6j$ symbols,
$g$ and $e$ point respectively ground and excited states.\\
\indent Formulas~(\ref{eq:MatrixElements}-\ref{eq:Unpert}) have been used to calculate frequency shift and modification
of intensity for corresponding transitions (for more details see~\cite{Leroy}). The frequency
shifts of atomic transitions \textit{2}, \textit{5}, \textit{8}, \textit{3}, \textit{6} and \textit{9}
versus magnetic $B$-field in
respect to initial position (the corresponding initial positions at $B = 0$
are indicated by the arrows) are presented by solid curves in Fig.~\ref{fig:FrShift}. The black
squares are the experimental results, i.e. frequency shifts of VSOPs labeled \textit{2},
\textit{5}, \textit{8}, \textit{3}, \textit{6}
and \textit{9} (numbers denote the corresponding transitions, see Fig.~\ref{fig:levels}). As it is seen the
theoretical model very well describes the observed results. In Fig.~\ref{fig:FrShift} are shown
only the atomic transitions (represented by
VSOPs) which are observed in the spectra (particularly, atomic transition
$F_g=1 \rightarrow F_e=0$ is omitted). It is interesting to note, that the frequency
of the transition labeled \textit{8} remains practically the same in
the range of $100 - 1100$~G, while $g$-factors for the ground and excited levels
are $-0.7$~MHz/G and $0.93$~MHz/G respectively. Thus, in the linear Zeeman effect (i.e.
when a frequency shift is proportional to $B$-field) one expects to detect
a shift of $\sim 1600$~MHz when $B = 1000$~G, while the shift is nearly zero.
Obviously, this is caused by the influence of the neighboring levels with the
selection rules presented by formula~(\ref{eq:SelRules}). Another interesting point
is that for transition labeled \textit{6} there is deviation from the linear Zeeman for
small field values of order of a few Gauss. Note, that atomic transition represented
by VSOP labeled \textit{2} can be a convenient one for an external magnetic
field measurement since it has constant frequency shift of $1.28$~MHz/G in the
whole region $1 - 1000$~G. Also, the probability of this atomic transition
is high enough and this is displayed by the large amplitude of VSOP
labeled \textit{2}.  The transition labeled \textit{9} indicates a higher constant frequency shift
that transition \textit{2}, but (see Fig.~\ref{fig:Lambda3}) intensity of transition \textit{9}
(displayed by
small amplitude of VSOP labeled \textit{9}) is essentially lower.\\
\begin{figure}
\begin{center}
\resizebox{0.6\columnwidth}{!}
{
\includegraphics{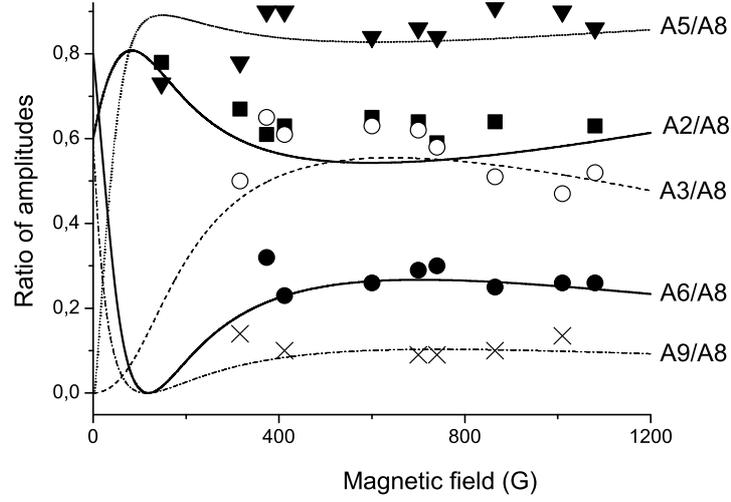}
}
\caption{The ratio of the amplitudes (i.e. ratio of the probabilities) $Ai/A8$
($i = 2,3,5,6, 9$)
in the case of $\pi$-excitation versus magnetic $B$-field:
experimental points are presented by squares,  hollow circles,  triangles,
filled circles, and crosses, correspondingly  and theory by   solid,
dashed, dotted,  solid, and dot-dashed lines, correspondingly.
It is easy to visually track behavior of the forbidden transitions \textit{3} and \textit{5}
since at $B = 0$ the probabilities are zero.}
\label{fig:Amplitudes}
\end{center}
\end{figure}
\indent Let's now consider the change in the probability of atomic transition
versus applied magnetic $B$-field. As it was demonstrated in
\cite{Hakhum2,Hakhum3,Leroy} the change in the probability (i.e. change of the dipole moment)
causes the change in the Rabi frequency of the laser radiation and, as a
consequence, the change in the efficiency of optical pumping process.
This is displayed as an increase or decrease in the corresponding VSOP
resonance amplitude presented on the spectra at Figs.~(\ref{fig:Lambda}-\ref{fig:Lambda3}). Note, that
in the experiment it is more convenient to measure the ratio of the VSOPs
amplitudes, \textit{A2}, \textit{A3}, \textit{A5}, \textit{A6}, \textit{A8} and \textit{A9} of the corresponding transitions
as a function of \textit{B}, since the absolute value of the VSOP
amplitude depends on the laser intensity, scanning time of an atomic
transition by a frequency of laser radiation, nano-cell temperature, \textit{etc}.
Consequently in Fig.~\ref{fig:Amplitudes} shown are experimental
(squares, hollow circles, triangles, filled circles, and crosses, correspondingly)
and theoretical (by   solid,  dashed, dotted,  solid, and dot-dashed cuves, correspondingly)
 ratios of the \textit{A2}, \textit{A3}, \textit{A5}, \textit{A6}, \textit{A9} divided by \textit{A8} (\textit{A8} is chosen since
the amplitude of VSOP denoted \textit{8} changes most slowly among the others).
Note, that coincidence of the experiment and the theory is not as good
as it is for the frequency shift. The explanation is as follows:
since the VSOP is located exactly at the atomic transitions, thus a
shift of the VSOP frequency displays exactly the shift of the atomic
transition. As to the VSOP amplitude, although, it linearly depends on
the probability of corresponding atomic transition, however some other
factors can also have a slight influence. However, if we consider VSOP resonances
 formed by the widely used SA technique, they are not useful for
 the above mentioned study since the ratio of the amplitudes of VSOP resonances
 completely doesn't match the ratio of the corresponding transition probabilities.
\section{Conclusion}
By experimental exploration of  the $\lambda $-Zeeman technique based on
Rb nano-cell use, we reveal for the first time (both experimentally
and theoretically) a strong modification of atomic transitions
probabilities versus magnetic $B$-field in the range of
$100 - 1100$~G, of the $^{87}$Rb, $D_2$ line $F_g=1 \rightarrow F_e=0, 1, 2, 3$ transitions,
including $F_g=1, m_F =0 \rightarrow F_e=1, m_F=0$ and $F_g=1, m_F =-1 \rightarrow F_e=3, m_F =-1$ transitions
(these transitions are "forbidden" when $B=0$). Note, that these two
forbidden transitions at $B > 150$~G are among the strongest atomic transitions in the detected spectra.\\
\indent It is shown that the observed frequency shifts of the $F_g=1 \rightarrow F_e=0, 1, 2, 3$ atomic
transitions in an external magnetic field $B>150$~G are very well described by the
developed theoretical model. Particularly, $F_g=1, m_F =+1 \rightarrow F_e=1, m_F =+1$ atomic
transition is a unique one since its frequency remains practically unchanged when
$B$ varies in the range of $100 - 1100$~G (this is evidence of another type of
non-linear Zeeman shift). For an external magnetic field measurement
the $F_g=1, m_F =-1 \rightarrow F_e=2, m_F =-1$ atomic transition can be a convenient one,
since it has a constant frequency shift of $1.28$~MHz/G in the whole
region of $1 - 1000$~G. Also, the probability of this atomic transition is
high enough and this is displayed by the large amplitude of the corresponding VSOP resonance.\\
\indent The experimental results are in a good agreement with the theoretical calculations.
It is worth to note, that the presented experimental results obtained with the help
of Rb nano-cell (i.e. $\lambda $-Zeeman technique) might be
obtained using expensive and complicated systems based either on cold
and trapped atoms or collimated several-meter long Rb
atomic beam propagating in vacuum conditions.\\
\indent Simple and robust $\lambda $-Zeeman technique can be successfully implemented also
for the study of the $D_1$ and $D_2$ lines of Na, K, Cs and other atoms.

\section{Acknowledgement}
The authors are grateful to A. Sarkisyan for his valuable participation
in fabrication of the NTC as well as to A. Papoyan, A. Sargsyan and
 A. Bagdasaryan for useful discussions. Research conducted in the scope of the International
 Associated Laboratory (CNRS-France \& SCS-Armenia)  IRMAS.

% BibTeX users please use
% \bibliographystyle{}
% \bibliography{}

\end{document}